\documentclass[twocolumn,showpacs,preprintnumbers,amsmath,amssymb]{revtex4}

\usepackage{epsfig}
\usepackage{dcolumn}
\usepackage{bm}
\usepackage{graphicx}
\usepackage{color}

\begin{document}


\title{Non-Equilibrium Dynamics of Correlated Electron Transfer in Molecular
  Chains}

\author{L.~M\"uhlbacher, J.~Ankerhold, and A.~Komnik}

\affiliation{Physikalisches Institut,
Albert-Ludwigs-Universit\"at, D-79104 Freiburg, Germany }

\date{\today}

\begin{abstract}
  The relaxation dynamics of correlated electron transport (ET) along molecular
  chains is studied based on a substantially improved numerically exact path
  integral Monte Carlo (PIMC) approach. As archetypical model we consider a
  Hubbard chain containing two interacting electrons coupled to a bosonic bath.
  For this generalization of the ubiquitous spin-boson model, the intricate
  interdependence of correlations and dissipation leads to non-Boltzmann
  thermal equilibrium distributions for many-body states. By mapping the
  multi-particle dynamics onto an isomorphic single particle motion this
  phenomenon is shown to be sensitive to the particle statistics and due to its
  robustness allows for new control schemes in designed quantum aggregates.
\end{abstract}




\maketitle

Recent progress in engineering and manufacturing of nanoscopic electronic
devices enabled to tackle the long standing problem of contacting single
molecules \cite{reed,weber,c60}. This remarkable achievement not only opens a
way for building the ultimate transistor on the molecular basis, but is also
expected to become an important tool for investigating electron transfer (ET),
which is one of the most fundamental processes in chemistry and biology
\cite{jortner}, the best known example being photosynthesis \cite{may}.
Likewise, in semiconductors charge transfer plays a key role and most recently
studied arrays of few electron quantum dots may be seen as artificial molecular
structures which, in contrast to their native counterparts, allow for a
completely controllable ET \cite{heinzel}.

The most successful paradigm for the description of ET is the donor--acceptor
model. In its simplest form it is represented by a two-site chain where a
single electron is transferred between the sites via tunneling hybridization.
However, irreversible transfer can only take place in presence of an
environment consisting e.g.\ of residual vibronic degrees of freedom or
electromagnetic fluctuations. In the past, the Caldeira--Leggett model
\cite{CLM} has been shown to capture the relevant features of dissipative ET,
where the electronic system is coupled to a bath of harmonic oscillators
\cite{weiss}. In case of only two sites and one single charge this is the
famous spin--boson model, which, together with its generalizations to
multi-site geometry, is a paradigmatic model on its own comprising e.g.\
quantum Brownian motion, Kondo physics, Luttinger liquids \cite{weiss}, and
atomic quantum dots \cite{zwerger}. Despite its simplicity the associated
phenomenology is extremely rich, showing different sorts of quantum phase
transitions \cite{dissipativeRMP}.

With the advent of fully tunable quantum structures a fundamental constraint
has to be relaxed though, namely, the restriction to single particle dynamics.
Multi-particle correlation effects may undoubtedly have a potential to
considerably alter the simple ET physics and open the door for new control
mechanisms \cite{park,liang}. Of course, correlated ET in a dissipative
reservoir is a formidable task, mainly due to the fact, that a quantum
reservoir leads to strong retardation effects. So far virtually all studies
made use of perturbative approaches focusing on either completely coherent
transfer with no dissipation \cite{wingreen,ferretti}, or completely sequential
transfer with very strong Coulomb repulsion \cite{petrov}, which effectively
reduces to a single particle ET problem.  The inclusion of both correlations
and dissipation has been attacked only recently in \cite{tornow}, where
equilibrium properties of a two site spinful system filled with two electrons
has been analyzed with the help of numerical renormalization group (RG)
technique. However, being intrinsically an equilibrium technique, this method
cannot reveal any information about dynamical properties of these systems. In
addition, finite temperature effects are usually quite difficult to access in
the framework of the RG. In this Letter we close this gap and, based on a
numerically exact path integral Monte Carlo (PIMC) scheme, we present results
for the real-time dynamics of two correlated fermions in a dissipative
environment. This model can be seen as the simplest generalization of the
single particle spin-boson model to include many-body effects. It may thus
reveal archetypical properties of correlated dissipative quantum dynamics.


The system is modeled by an open Hubbard chain with $N$ sites of spacing $a$
\begin{eqnarray} \label{H00}
 H_S &=& \sum_{i=1, \sigma=\uparrow,\downarrow}^{N} \, E_i \, d^\dag_{i \sigma}
 \, d_{i \sigma} + \frac{U_i}{2} \, d^\dag_{i \sigma} d_{i \sigma} \,
  d^\dag_{i -\sigma} d_{i -\sigma}
\nonumber \\
 && {} + \sum_{i=1, \sigma=\uparrow,\downarrow}^{N-1}
 \Delta_i \left( d^\dag_{i \sigma} \, d_{i+1 \sigma} +
 h.c. \right) \, ,
\end{eqnarray}
where $d_{i \sigma}$ are annihilation operators for electrons with spin
$\sigma$ on the site $i$. $E_i$ are the bare energies of the levels, $U_i$ the
corresponding interaction strengths and $\Delta_i$ are the tunneling matrix
elements. The interaction with the bosonic bath given by the Hamiltonian $H_B =
\sum_\alpha \left({P_\alpha^2 / 2m_\alpha} + m_\alpha \omega_\alpha^2
  X_\alpha/2 \right)$ is accomplished via a standard dipole coupling
\begin{eqnarray} \label{couplingterm}
 H_I = - a {\cal P} \sum_\alpha c_\alpha X_\alpha + a^2{\cal P}^2\sum_\alpha
 {c_\alpha^2 \over 2m_\alpha \omega_\alpha^2} \, ,
\end{eqnarray}
where
\begin{eqnarray} {\cal P} = \sum_{i=1, \sigma}^N \, [i -
(N+1)/2] \, d_{i \sigma}^\dag d_{i \sigma}
\end{eqnarray}
is the polarization operator of the Hubbard chain.

We are interested in the reduced dynamics of the fermionic system determined by
the density operator $\rho(t)={\rm Tr}_B\{ \exp(-i H t/\hbar)\, W(0)\, \exp(i H
t/\hbar)\}$, where $W(0)$ specifies the initial state of the compound.  Here
the trace over the bosonic bath degrees of freedom can be done exactly within
the path integral approach by representing the density operator in the site
representation of the chain in terms of a pseudo-spin-$(N-1)/2$ operator
$\vec{S}^\sigma$.  Accordingly, the eigenvalues of $S_z^\sigma$ represent the
position of the respective electron along the chain, i.e.\ $S_z^\sigma
|s^\sigma\rangle = s^\sigma |s^\sigma\rangle$ ($s^\sigma=-S,\ldots, S$), and
$|s^\sigma\rangle$ denote a localized single-electron state. In the rest of the
Letter we focus on a chain filled with two electrons of opposite spins with
$\sigma={\uparrow, \downarrow}$. Then, the corresponding many-body basis
$\{|s^\uparrow, s^\downarrow\rangle\}$ follows from properly anti-symmetrized
single particle states. The relevant dynamical observables are the populations
$P_{\lambda}(t)={\rm Tr}_S\{\lambda \rho(t)\}$ where $\lambda$ denotes a
projection operator either onto a certain site and spin or onto a certain
many-body state.  Initially, the system is prepared in a specific many-body
state with the bosonic reservoir equilibrated to that state.

This way, the reduced density operator is expressed as a double path integral
along a Keldysh contour with forward $s^\sigma$ and backward $\tilde{s}^\sigma$
paths. The impact of the dissipative environment appears as an influence
functional introducing arbitrarily long-ranged interactions in time between the
paths. It is convenient to switch to the combinations
${\eta}^\sigma=s^\sigma+\tilde{s}^\sigma$ and
${\xi}^\sigma=s^\sigma-\tilde{s}^\sigma$ so that one arrives at the exact
expression
\begin{equation}
P_\lambda(t) = \oint\!{\cal D}\vec{\eta} \oint\!{\cal
D}\vec{\xi}\; \lambda(\vec{\eta},\vec{\xi}) {\cal
A}[\vec{\eta},\vec{\xi}]\, \exp\!\left( -
 \Phi[\vec{\eta},\vec{\xi}] \right) \label{pops}\; ,
\end{equation}
where $\vec{\eta}=(\eta^\uparrow,\eta^\downarrow)$ and $
\vec{\xi}=(\xi^\uparrow,\xi^\downarrow)$. Here, ${\cal A}$ is the bare action
factor in absence of a reservoir and the influence functional reads
\begin{eqnarray}
\Phi[\vec{\eta},\vec{\xi}] &=& \int_0^t ds \int_0^s du
\left\{\left[\vec{\xi}(s) \cdot \vec{e}\right] L'(s-u)
\left[\vec{\xi}(u) \cdot \vec{e}\right]\right.\nonumber\\
&&\left.\hspace{1.5cm} +i \left[\vec{\xi}(s) \cdot
\vec{e}\right] L''(s-u)
\Big[\vec{\eta}(u) \cdot \vec{e}\Big]\right\}\nonumber\\
&& +i \frac{\mu}{2}\int_0^t ds \left[ \vec{\xi}(s) \cdot
\vec{e}\right]\Big[\vec{\eta}(s) \cdot
\vec{e}\Big]\nonumber
\end{eqnarray}
with $\vec{e}=(1,1)$. The kernel $L(t)=L'(t)+i L''(t)$ is related to the
force-force auto-correlation function of the bath and is completely determined
by the spectral density of its modes. Further, $\mu=\lim_{\hbar\beta\to 0}
\hbar\beta L(0)$. The immediate advantage of the representation (\ref{pops}) is
that formally it looks like the dynamics of a single particle in a
two-dimensional plane. We shall discuss this point in more detail below.  Note
further, that even in absence of a direct Coulomb interaction the two charges
are correlated due to the coupling to the heat bath.

An analytical treatment of the expression (\ref{pops}) is in general not
feasible, mainly due to the retardation in the influence functional which grows
with decreasing temperature. In this situation PIMC methods have been shown to
be very powerful numerically exact means to explore the non-perturbative range.
Our starting point is a PIMC scheme \cite{lothar1} which exploits the linear
dependence of $\Phi[\vec{\eta},\vec{\xi}]$ on the quasi-classical paths
$\vec{\eta}(t)$, so that in (\ref{pops}) the corresponding summations can be
expressed as a series of simple matrix multiplications and therefore be carried
out explicitly. These matrices include the bare short time propagators of the
system and phase factors stemming from the $\vec{\eta}$-dependent part of
$\Phi$.  Consequently, the number of MC variables is reduced by a factor of
two, which leads to a significant soothing of the dynamical sign problem
\cite{egger}. Previous simulations were restricted to single particle transport
on up to three sites and for shorter times only, because the standard MC-weight
requires to perform the full series of matrix multiplications for each MC-step
due to the retardation of the $\vec{\eta}$-dependent phase factor. A
breakthrough is gained by neglecting the long-time retardations during the
propagation of the MC trajectory.  Since they are fully accounted for during
the final accumulation process, the numerical exactness of the MC scheme is
{\em not} impaired. This way, one achieves a strong decoupling of
quasi-classical ($\vec{\eta}$) and quantum ($\vec{\xi}$) coordinates, which
allows to calculate and store products of short time propagators independent of
the MC-sampling. Eventually a speed-up with respect to the original method
\cite{egger} by a factor of about 100 is gained.

\begin{figure}
\vspace*{0.7cm}
\epsfig{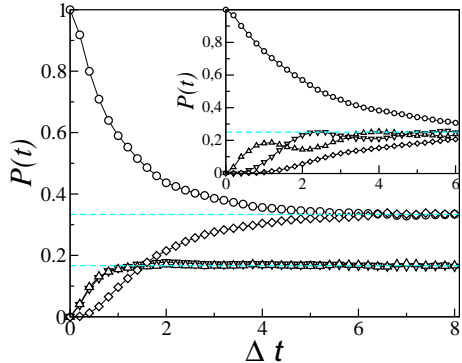}
\caption[]{\label{fig1} Relaxation dynamics of two correlated electrons in a
  degenerate two-site systems according to Fig.~\ref{figX}a. Shown are the
  populations for the states $|\!\downarrow\uparrow,0\rangle$ (circles),
  $|\!\downarrow,\uparrow\rangle$ (triangles up),
  $|\!\uparrow,\downarrow\rangle$ (triangles down), and
  $|0,\!\downarrow\uparrow\rangle$ (diamonds) for a bath with ohmic spectral
  density and exponential cutoff at $\omega_c = 5\Delta$; the damping strength
  is $\alpha=1/4$ and temperature $\hbar\beta\Delta=0.1$. Dashed lines
  represent theoretical equilibrium values. Inset: Dynamics for a system
  according to Fig.~\ref{figX}b. Error bars (vertical lines) are typically of
  the size of the symbols.}
\vspace*{0.0cm}
\end{figure}

The simplest non-trivial system which is able to accommodate two electrons and
thus the simplest many-body generalization of the spin-boson model is a
two-site chain. The proper representation of a corresponding many-body basis is
given by
\begin{equation} \label{localized states}
\mathcal{B} = \{ |\!\downarrow \uparrow, 0\rangle\,,\,
|\!\downarrow, \uparrow\rangle\,,\,
  |\!\uparrow, \downarrow\rangle\,,\, |0, \downarrow \uparrow\rangle\} \,
  ,
\end{equation}
where the left(right) argument of the ket represents the spin occupation of
site $-1/2(+1/2)$. For these localized spin states the relevant observables are
the site populations and the populations ${P}_\psi$ occupying one of the states
$|\psi\rangle \in \mathcal{B}$. In particular, the latter ones give detailed
insight into the intimate relation between correlations and dissipation during
the relaxation from an initial non-equilibrium state. For very long times, the
expectation is that the system reaches a thermal equilibrium, where all four
many-body states are populated according to a Boltzmann distribution. Thus, in
case of vanishing Coulomb interaction $U_i=0$ and energetically completely
degenerated sites $E_i=0$, one anticipates, no matter how the system is
prepared at $t=0$, a final probability of 1/4 for each state in (\ref{localized
  states}), whereas both sites of the chain have to be half-filled. While the
latter expectation turns out to be true, remarkably, the former one cannot be
confirmed by our simulations (see Fig.~\ref{fig1}).
Not only do the populations ${P}_\psi(t)$ saturate at different values, but
these also depend on the initial preparation. This at first startling dynamics
leading to a non-Boltzmann distributed steady state in the many-body basis
(\ref{localized states}), can be understood in terms of \emph{decoherence free
  states} (DFS) \cite{QuantumComputingBook}. These states are simultaneous
eigenstates of $H_S$ and of the polarization operator ${\cal P}$.  Accordingly,
after being prepared in such a state the system stays there forever and no
relaxation into the equilibrium can occur. While DFS have been studied for very
weakly damped systems in the context of quantum information processing, where
the generic basis is the eigenstate basis of $H_S$, in strongly condensed phase
systems the site representation is the proper one. Hence, for correlated ET the
impact of DFS onto {\em transport} properties is important and leads to
completely new and unexpected phenomena. In the case considered here, there is
only one such state, namely
\[
|\phi_1\rangle = (|\!\downarrow,\uparrow \rangle -
|\!\uparrow,\downarrow \rangle)/\sqrt{2} \, ,
\]
which corresponds to zero dipole moment and apparently is a delocalized spin
state. An alternative basis set of the electronic Hilbert space is thus given
by
\[
\{|\phi_i\rangle\}_{i \le 4} = \left\{|\phi_1\rangle,
\frac{1}{\sqrt{2}}(|\!\downarrow,\uparrow \rangle +
|\!\uparrow,\downarrow \rangle) , |\!\downarrow \uparrow,
0\rangle\,,\, |0,\!\downarrow \uparrow\rangle \right\}
\]
consisting of both delocalized ($|\phi_1\rangle$, $|\phi_2\rangle$) and
localized ($|\phi_3\rangle$, $|\phi_4\rangle$) spin states. Obviously,
$|\phi_1\rangle$ does not participate in the relaxation process, such that
equilibration is restricted to the sub-space spanned by $|\phi_l\rangle,
l=2,3,4$, leading for an energetically degenerate system to
\begin{eqnarray} \label{wichtig}
 {P}_{\phi_{l}}(t\to\infty) = [1-{P}_{\phi_1}(t=0)]/3\ ,\ l=2,3,4
\end{eqnarray}
Now, due to ${P}_{\downarrow\uparrow,0} = P_{\phi_3}$,
${P}_{0,\downarrow\uparrow} = P_{\phi_4}$, ${P}_{\downarrow,\uparrow}
+{P}_{\uparrow,\downarrow} = P_{\phi_2} + P_{\phi_1}$, and
${P}^\infty_{\downarrow,\uparrow}={P}^\infty_{\uparrow,\downarrow}$ for
$t\to\infty$ due to symmetry, one easily understands from (\ref{wichtig}) both
the non-Boltzmann distribution of the localized states as well as their
dependence on the initial preparation. Imagine that initially the electronic
sub-system is prepared in one of the states in (\ref{localized states}), which
experimentally is the typical situation. Then, for the non-equilibrium dynamics
two different scenarios are possible: If both electrons initially occupy the
same site, i.e.~${P}_{\downarrow\uparrow,0}(0) = 1$ or
${P}_{0,\downarrow\uparrow}(0) = 1$, one always has $P_{\phi_1}(0) = 0$ so that
the final equilibrium distributions read
${P}_{\downarrow\uparrow,0}^\infty = {P}_{0,\downarrow\uparrow}^\infty = 1/3$
and ${P}_{\downarrow,\uparrow}^\infty = {P}_{\uparrow,\downarrow}^\infty =
1/6$.  If, however, both sites are initially occupied,
i.e.~${P}_{\downarrow,\uparrow}(0) = 1$ or ${P}_{\uparrow,\downarrow}(0) = 1$,
so that $P_{\phi_1}(0) = 1/2$, the final distributions become
${P}_{\downarrow\uparrow,0}^\infty = {P}_{0,\downarrow\uparrow}^\infty = 1/6$
and ${P}_{\downarrow,\uparrow}^\infty = {P}_{\uparrow,\downarrow}^\infty =
1/3$.  Therefore, the system, after having been prepared in one of the states
(\ref{localized states}), always {\it remembers}, whether this state had a
vanishing or non-vanishing dipole moment. This phenomenon crucially depends on
the symmetry of the operator mediating the interaction with the bosonic baths.
However, since the dipole coupling specified in (\ref{couplingterm}) is generic
for a variety of native and artificial molecular structures, the above dynamics
can be expected to be typical. In most native molecules ET is triggered by
photo-excitation from an electronic ground state \cite{may} so that correlated
ET is difficult to observe.  In contrast, e.g.\ in synthetic aggregates, which
are very weakly coupled to external leads \cite{park,liang}, or in arrays of
quantum dots \cite{heinzel}, the excess charge can be better controlled.  Then,
starting e.g.\ from a totally unpolarized ensemble, after equilibration only
$1/3$ of its constituents would be polarized and $2/3$ of them unpolarized.
This is an example of a non-Boltzmann type of thermal distribution, which for
the generic form of system-bath coupling is a signature of: (i) the topology of
the tunneling couplings in $H_S$ [cf.\ (\ref{H00})] and (ii) Fermi-Dirac
statistics.  Let us analyze these two points in more detail.
\begin{figure}
\vspace*{0.7cm}
\epsfig{file=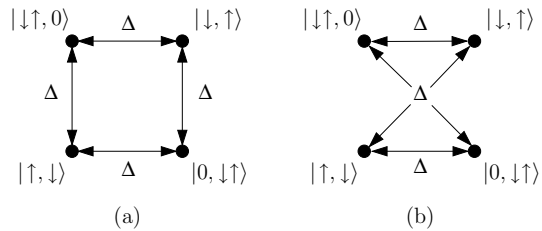,width=7cm}
\caption[]{\label{figX} Two-site Hubbard chain with standard (a) and spin
  flip/two particle tunneling (b) couplings.} \vspace*{0.0cm}
\end{figure}

\begin{figure}
\vspace*{0.7cm}
\epsfig{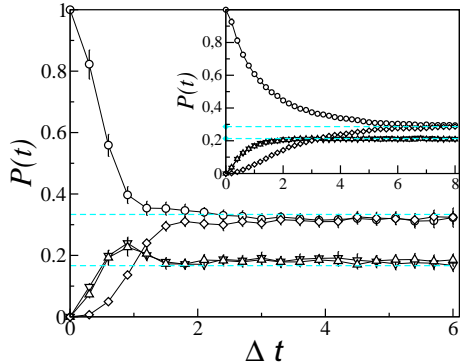}
\caption[]{\label{fig2} Same as in Fig.~\ref{fig1}, but for $\hbar\beta\Delta =
  0.5$. Inset: Dynamics for a system with $\hbar\beta\Delta = 0.1$, Coulomb
  interaction $U_1 = U_2 = 4\hbar\Delta$, and $E_1 = E_2 = 0$, i.e.~$\langle
  \downarrow\uparrow, 0| H_S|\downarrow \uparrow, 0\rangle= \langle
  0,\downarrow \uparrow|H_S|0,\downarrow \uparrow\rangle= 4\hbar\Delta, \langle
  \downarrow,\uparrow| H_S|\downarrow, \uparrow \rangle= \langle
  \uparrow,\downarrow| H_S|\uparrow,\downarrow \rangle = 0$.}
\vspace*{0.0cm}
\end{figure}

To address (i), one first realizes that the existence of a common eigenstate
$|\phi_1\rangle$ of $H_0$ and $\mathcal{P}$ is completely independent of the
electronic energies $E_i$, the Coulomb interaction strengths $U_i$, temperature
and the spectral density of the bath modes.  Accordingly, the described
phenomena can also be observed for non--degenerate systems ($E_1 \neq E_2\neq
0$ and/or $U_1 \neq U_2 \neq 0$). In this case, one even arrives at the
counter-intuitive situation that in thermal equilibrium the occupations of
energetically higher lying states {\em exceed} those of energetically lower
lying ones (cf.~inset in Fig.~\ref{fig2}). Moreover, effects tied to the DFS
also survive the coherent--incoherent transition, its typical signature being
the onset of oscillations in the population dynamics (cf.~Fig.~\ref{fig2}).
Thus, they prevail in two distinct dynamical regimes.
However, DFS crucially depend on the topology of the tunneling couplings in
$H_S$. They always exist for isotropic couplings, but can be easily destroyed
e.g.\ by appropriately adding or removing couplings between the localized
states (\ref{localized states}). For example, interchanging the coupling
according to Fig.~\ref{figX}b so that spin flips and two-particle tunneling are
allowed, the Boltzmann equilibrium is restored (see inset in Fig.~\ref{fig1}).
\begin{figure}
\vspace*{0.7cm} \epsfig{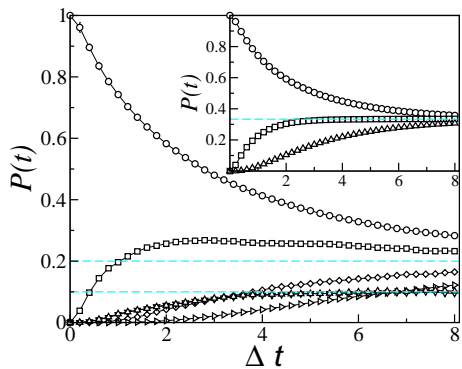}
\caption[]{\label{fig3} Same as in Fig.~\ref{fig1} but for two identical bosons
  on a chain with $N=3$. Shown are the populations for states
  $|n_{-1},n_0,n_1\rangle$ with occupation $n_i$ on site $i$: $|2,0,0\rangle$
  (circles), $|1,1,0\rangle$ (squares), $|1,0,1\rangle$ (triangles up),
  $|0,2,0\rangle$ (triangles down), $|0,1,1\rangle$ (diamonds), and
  $|0,0,2\rangle$ (triangles right). Inset: Dynamics for $N=2$ with states
  $|n_{-1/2},n_{1/2}\rangle$; $|2,0\rangle$ (circles), $|1,1\rangle$(squares),
  $|0,2\rangle$ (triangles).} \vspace*{0.0cm}
\end{figure}

To discuss (ii), one easily deduces a mapping of the dissipative dynamics of
two correlated fermions on two sites onto a single particle motion along the
edges of a two-dimensional square, spanned by the many body states
(\ref{localized states}), see Fig.~\ref{figX}a. This mapping has two immediate
consequences that will be studied in details elsewhere: First, it allows to
adapt perturbative approaches like the NIBA and the NICA \cite{weiss} developed
for ordinary spin-boson models to capture also correlated many-body dynamics;
on the other hand, the dynamics along any $N$-site chain with an arbitrary
number of electrons can be mapped onto a dissipative single particle ET on a
higher-dimensional square \cite{lothar2}. For indistinguishable bosons such a
mapping is also found leading though to a triangular lattice. Then one derives
from the geometry of the mapping that for a Hubbard chain only one DFS exists
whereas for the similar bosonic model such states are only possible for
\textit{odd} $N$. This is illustrated in Fig.~\ref{fig3} for chains with $N=3$
and $N=2$. In the former case one has final populations of 1/5 vs.\ 1/10, while
in the latter one they all saturate at 1/3. Note that the $N=3$ fermionic
system exhibits smaller equilibrium values of 1/8 vs.\ 1/12 due to the larger
number of accessible many-body states. Of course, DFS \emph{never} exist in
one-dimensional single particle systems.

To summarize, we have developed a PIMC technique which opens the door for a
reliable modeling of the nonequilibrium dynamics of correlated ET even on
longer time scales. As a remarkable many-body effect, related to the existence
of a DFS, a non-Boltzmann steady state was predicted. A mapping onto an
isomorphic single particle model reveals its sensitivity to the particle
statistics.  Experimentally, this phenomenon may lead to new control schemes
for fermionic or bosonic transport in tight-binding systems like synthesized
molecular chains, quantum dot arrays, or optical lattices.

We wish to thank H.~Grabert for many inspiring discussions. This work was
supported by the Landesstiftung Baden-W\"urttemberg, the DFG, and the EU.

\bibliography{PRLDissChain_short}

\begin{thebibliography}{20}
\expandafter\ifx\csname natexlab\endcsname\relax\def\natexlab#1{#1}\fi
\expandafter\ifx\csname bibnamefont\endcsname\relax
  \def\bibnamefont#1{#1}\fi
\expandafter\ifx\csname bibfnamefont\endcsname\relax
  \def\bibfnamefont#1{#1}\fi
\expandafter\ifx\csname citenamefont\endcsname\relax
  \def\citenamefont#1{#1}\fi
\expandafter\ifx\csname url\endcsname\relax
  \def\url#1{\texttt{#1}}\fi
\expandafter\ifx\csname urlprefix\endcsname\relax\def\urlprefix{URL }\fi
\providecommand{\bibinfo}[2]{#2}
\providecommand{\eprint}[2][]{\url{#2}}

\bibitem[{\citenamefont{Reed et~al.}(1997)\citenamefont{Reed, Zhou, Muller
  et~al.}}]{reed}
\bibinfo{author}{\bibfnamefont{M.~A.} \bibnamefont{Reed}},
  \bibinfo{author}{\bibfnamefont{C.}~\bibnamefont{Zhou}},
  \bibinfo{author}{\bibfnamefont{C.~J.} \bibnamefont{Muller}},
  \bibnamefont{et~al.}, \bibinfo{journal}{Science}
  \textbf{\bibinfo{volume}{278}}, \bibinfo{pages}{252} (\bibinfo{year}{1997}).

\bibitem[{\citenamefont{Reichert et~al.}(2002)\citenamefont{Reichert, Ochs,
  Beckmann et~al.}}]{weber}
\bibinfo{author}{\bibfnamefont{J.}~\bibnamefont{Reichert}},
  \bibinfo{author}{\bibfnamefont{R.}~\bibnamefont{Ochs}},
  \bibinfo{author}{\bibfnamefont{D.}~\bibnamefont{Beckmann}},
  \bibnamefont{et~al.}, \bibinfo{journal}{Phys. Rev. Lett.}
  \textbf{\bibinfo{volume}{88}}, \bibinfo{pages}{176804}
  (\bibinfo{year}{2002}).

\bibitem[{\citenamefont{Park et~al.}(2000)\citenamefont{Park, Park, Lim
  et~al.}}]{c60}
\bibinfo{author}{\bibfnamefont{H.}~\bibnamefont{Park}},
  \bibinfo{author}{\bibfnamefont{J.}~\bibnamefont{Park}},
  \bibinfo{author}{\bibfnamefont{A.~K.~L.} \bibnamefont{Lim}},
  \bibnamefont{et~al.}, \bibinfo{journal}{Nature}
  \textbf{\bibinfo{volume}{407}}, \bibinfo{pages}{57} (\bibinfo{year}{2000}).

\bibitem[{\citenamefont{Jortner and Bixon}(1999)}]{jortner}
\bibinfo{editor}{\bibfnamefont{J.}~\bibnamefont{Jortner}} \bibnamefont{and}
  \bibinfo{editor}{\bibfnamefont{M.}~\bibnamefont{Bixon}}, eds.,
  \emph{\bibinfo{title}{Electron Transfer from isolated Molecules to
  Biomolecules}}, vol. \bibinfo{volume}{106/107} of \emph{\bibinfo{series}{Adv.
  Chem. Phys.}} (\bibinfo{year}{1999}).

\bibitem[{\citenamefont{May and K\"uhn}(2004)}]{may}
\bibinfo{author}{\bibfnamefont{V.}~\bibnamefont{May}} \bibnamefont{and}
  \bibinfo{author}{\bibfnamefont{O.}~\bibnamefont{K\"uhn}},
  \emph{\bibinfo{title}{Charge and Energy Transfer Dynamics in Molecular
  Systems}} (\bibinfo{publisher}{Wiley--VCH, Weinheim}, \bibinfo{year}{2004}).

\bibitem[{\citenamefont{Heinzel}(2003)}]{heinzel}
\bibinfo{author}{\bibfnamefont{T.}~\bibnamefont{Heinzel}},
  \emph{\bibinfo{title}{Mesoscopic Electronics in Solid State Nanostructures}}
  (\bibinfo{publisher}{Wiley--VCH, Weinheim}, \bibinfo{year}{2003}).

\bibitem[{\citenamefont{Caldeira and Leggett}(1983)}]{CLM}
\bibinfo{author}{\bibfnamefont{A.~O.} \bibnamefont{Caldeira}} \bibnamefont{and}
  \bibinfo{author}{\bibfnamefont{A.~J.} \bibnamefont{Leggett}},
  \bibinfo{journal}{Physica A} \textbf{\bibinfo{volume}{121}},
  \bibinfo{pages}{587} (\bibinfo{year}{1983}).

\bibitem[{\citenamefont{Weiss}(1999)}]{weiss}
\bibinfo{author}{\bibfnamefont{U.}~\bibnamefont{Weiss}},
  \emph{\bibinfo{title}{Quantum dissipative systems}}
  (\bibinfo{publisher}{World Scientific}, \bibinfo{year}{1999}),
  \bibinfo{note}{and references therein}.

\bibitem[{\citenamefont{Recati et~al.}(2005)\citenamefont{Recati, Fedichev,
  Zwerger et~al.}}]{zwerger}
\bibinfo{author}{\bibfnamefont{A.}~\bibnamefont{Recati}},
  \bibinfo{author}{\bibfnamefont{P.~O.} \bibnamefont{Fedichev}},
  \bibinfo{author}{\bibfnamefont{W.}~\bibnamefont{Zwerger}},
  \bibnamefont{et~al.}, \bibinfo{journal}{Phys.~Rev.~Lett.}
  \textbf{\bibinfo{volume}{94}}, \bibinfo{pages}{040404}
  (\bibinfo{year}{2005}).

\bibitem[{\citenamefont{Leggett et~al.}(1987)\citenamefont{Leggett,
  Chakravarty, Dorsey et~al.}}]{dissipativeRMP}
\bibinfo{author}{\bibfnamefont{A.~J.} \bibnamefont{Leggett}},
  \bibinfo{author}{\bibfnamefont{S.}~\bibnamefont{Chakravarty}},
  \bibinfo{author}{\bibfnamefont{A.~T.} \bibnamefont{Dorsey}},
  \bibnamefont{et~al.}, \bibinfo{journal}{Rev.~Mod.~Phys.}
  \textbf{\bibinfo{volume}{59}}, \bibinfo{pages}{1} (\bibinfo{year}{1987}).

\bibitem[{\citenamefont{Park et~al.}(2002)\citenamefont{Park, Pasupathy,
  Goldsmith et~al.}}]{park}
\bibinfo{author}{\bibfnamefont{J.}~\bibnamefont{Park}},
  \bibinfo{author}{\bibfnamefont{A.~N.} \bibnamefont{Pasupathy}},
  \bibinfo{author}{\bibfnamefont{J.~I.} \bibnamefont{Goldsmith}},
  \bibnamefont{et~al.}, \bibinfo{journal}{Nature}
  \textbf{\bibinfo{volume}{417}}, \bibinfo{pages}{722} (\bibinfo{year}{2002}).

\bibitem[{\citenamefont{Liang et~al.}(2002)\citenamefont{Liang, Shores,
  Bockrath et~al.}}]{liang}
\bibinfo{author}{\bibfnamefont{W.~J.} \bibnamefont{Liang}},
  \bibinfo{author}{\bibfnamefont{M.~P.} \bibnamefont{Shores}},
  \bibinfo{author}{\bibfnamefont{M.}~\bibnamefont{Bockrath}},
  \bibnamefont{et~al.}, \bibinfo{journal}{Nature}
  \textbf{\bibinfo{volume}{417}}, \bibinfo{pages}{725} (\bibinfo{year}{2002}).

\bibitem[{\citenamefont{Meir and Wingreen}(1992)}]{wingreen}
\bibinfo{author}{\bibfnamefont{Y.}~\bibnamefont{Meir}} \bibnamefont{and}
  \bibinfo{author}{\bibfnamefont{N.}~\bibnamefont{Wingreen}},
  \bibinfo{journal}{Phys.~Rev.~Lett.} \textbf{\bibinfo{volume}{68}},
  \bibinfo{pages}{2512} (\bibinfo{year}{1992}).

\bibitem[{\citenamefont{Ferretti et~al.}(2005)\citenamefont{Ferretti,
  Calzolari, Felice et~al.}}]{ferretti}
\bibinfo{author}{\bibfnamefont{A.}~\bibnamefont{Ferretti}},
  \bibinfo{author}{\bibfnamefont{A.}~\bibnamefont{Calzolari}},
  \bibinfo{author}{\bibfnamefont{R.~D.} \bibnamefont{Felice}},
  \bibnamefont{et~al.}, \bibinfo{journal}{Phys. Rev. Lett.}
  \textbf{\bibinfo{volume}{94}}, \bibinfo{pages}{116802}
  (\bibinfo{year}{2005}).

\bibitem[{\citenamefont{Petrov and H\"anggi}(2001)}]{petrov}
\bibinfo{author}{\bibfnamefont{E.~G.} \bibnamefont{Petrov}} \bibnamefont{and}
  \bibinfo{author}{\bibfnamefont{P.}~\bibnamefont{H\"anggi}},
  \bibinfo{journal}{Phys. Rev. Lett.} \textbf{\bibinfo{volume}{86}},
  \bibinfo{pages}{2862} (\bibinfo{year}{2001}).

\bibitem[{\citenamefont{Tornow et~al.}(2005)\citenamefont{Tornow, Tong, and
  Bulla}}]{tornow}
\bibinfo{author}{\bibfnamefont{S.}~\bibnamefont{Tornow}},
  \bibinfo{author}{\bibfnamefont{N.-H.} \bibnamefont{Tong}}, \bibnamefont{and}
  \bibinfo{author}{\bibfnamefont{R.}~\bibnamefont{Bulla}},
  \bibinfo{journal}{cond-mat/0502276}  (\bibinfo{year}{2005}).

\bibitem[{\citenamefont{M\"uhlbacher and Ankerhold}(2005)}]{lothar1}
\bibinfo{author}{\bibfnamefont{L.}~\bibnamefont{M\"uhlbacher}}
  \bibnamefont{and}
  \bibinfo{author}{\bibfnamefont{J.}~\bibnamefont{Ankerhold}},
  \bibinfo{journal}{J.~Chem.~Phys.} \textbf{\bibinfo{volume}{122}},
  \bibinfo{pages}{184715} (\bibinfo{year}{2005}).

\bibitem[{\citenamefont{Egger and Mak}(1994)}]{egger}
\bibinfo{author}{\bibfnamefont{R.}~\bibnamefont{Egger}} \bibnamefont{and}
  \bibinfo{author}{\bibfnamefont{C.}~\bibnamefont{Mak}},
  \bibinfo{journal}{Phys. Rev. B} \textbf{\bibinfo{volume}{50}},
  \bibinfo{pages}{15210} (\bibinfo{year}{1994}).

\bibitem[{\citenamefont{Nielsen and Huang}(2000)}]{QuantumComputingBook}
\bibinfo{author}{\bibfnamefont{M.~A.} \bibnamefont{Nielsen}} \bibnamefont{and}
  \bibinfo{author}{\bibfnamefont{I.~L.} \bibnamefont{Huang}},
  \emph{\bibinfo{title}{Quantum Computation and Quantum Information}}
  (\bibinfo{publisher}{Cambridge}, \bibinfo{year}{2000}).

\bibitem[{\citenamefont{M\"uhlbacher et~al.}(2005)\citenamefont{M\"uhlbacher,
  Escher, and Ankerhold}}]{lothar2}
\bibinfo{author}{\bibfnamefont{L.}~\bibnamefont{M\"uhlbacher}},
  \bibinfo{author}{\bibfnamefont{C.}~\bibnamefont{Escher}}, \bibnamefont{and}
  \bibinfo{author}{\bibfnamefont{J.}~\bibnamefont{Ankerhold}},
  \bibinfo{journal}{J.~Chem.~Phys.}  (\bibinfo{year}{2005}), \bibinfo{note}{in
  press}.

\end{thebibliography}

\end{document}